\documentstyle{article}
\begin{document}
\title{\huge Transition between Atomic and Molecular Hydrogen in the Galaxy:\\
Vertical Variation of the Molecular Fraction }
\author{Kouitiro Imamura and Yoshiaki Sofue\\
Institute of Astronomy, University of Tokyo, Mitaka, Tokyo 181, Japan} 
%\date{December 1995}
\date{}
\maketitle
%\vspace{5cm}
\vfill
\begin{abstract}
\large

We derive radial and vertical distributions of HI and H$_2$ gas densities in 
our Galaxy by using the terminal velocity method. We calculate the molecular 
fraction ($f_{\rm{mol}}$) defined as the ratio of the molecular hydrogen to 
total hydrogen gas density at galactic longitude $l= 33^\circ \sim 64^\circ$ 
and galactic latitude $b=-2^\circ \sim +2^\circ$. The thickness of the 
molecular dominant region $(f_{\rm{mol}} \geq 0.8)$ is approximately 
constant ($109\pm 12$ pc) at galactocentric distance $R\simeq$4.7$-$7.2 kpc. 
The molecular fraction decreases suddenly at a critical height from the 
galactic plane, below which the gas disk is almost totally molecular, 
while it is almost atomic beyond this height. We show that the 
vertical $f_{\rm{mol}}$ variation can be reproduced by a model which takes 
into account the phase transition between HI and H$_2$ gases in the 
interstellar matter.
\end{abstract}

\newpage\large

\section{Introduction}
The major constituents of the interstellar matter (ISM) in the Galaxy  
are the neutral atomic (HI) and molecular (H$_2$) hydrogen gases. 
The distributions of the HI and H$_2$ gases have been obtained through 
observations of the 21-cm and CO-line emissions, respectively. 
It is known that the HI gas is distributed in the outer region of the Galaxy, 
while H$_2$ gas is more concentrated toward the galactic center
(e.g., Burton 1988; Combes 1991). However, the mutual relation of the 
distributions of HI and H$_2$ gases in the Galaxy, which will be deeply 
coupled with the transition between these two phases of the ISM, has been 
not yet investigated in detail.

Elmegreen (1993) has proposed a phase-transition theory of the HI and H$_2$ 
gases, in which the molecular fraction ($f_{\rm{mol}}$), as defined by the 
ratio of the H$_2$ gas mass density to the total hydrogen gas density, is 
determined by the interstellar pressure, UV radiation intensity and the 
metal abundance. Sofue {\it et al.} (1995) and Honma {\it et al.} (1995) 
have studied the radial variation of $f_{\rm{mol}}$ in several galaxies, 
and found that the molecular fraction decreases suddenly at a certain 
galactocentric distance, which they called the molecular front. 
They further modeled the molecular front phenomenon based on the 
phase-transition theory of the HI and H$_2$ gases in the ISM.

Since the Milky Way Galaxy has been observed extensively both in the HI and 
CO lines at much higher linear resolution than external galaxies, we may be 
able to study the molecular front phenomenon in the Galaxy not only in the 
radial direction but also in the perpendicular direction to the 
galactic plane. In this paper we analyze latitude-velocity ($b-v$) 
diagrams of the HI- and CO-line emissions of the Galaxy by applying the 
terminal velocity method, and derive the HI and H$_2$ gas distributions 
in the $(R,z)$ plane, where $R$ is the galactocentric distance and $z$ 
is the height from the galactic plane. We further derive the distribution 
of the molecular fraction in the ($R,z$) plane, and investigate the hydrogen 
phase transition in the $z$ direction. We also model the result based on the 
phase-transition theory of ISM.

\section{Data and Procedure}
\subsection{Data}
We use the HI survey of Bania and Lockman (1984) observed with the
Arecibo 304-m radio telescope. The HI data are in the forming $b-v$
diagrams, and maps are sampled at intervals of $\Delta b = 2'$ with a
half-power beamwidth of $4'$, and the latitude coverage is $\pm 3^\circ$.
We also use the CO survey of Knapp {\it et al.} (1985) observed with the
7-m telescope at the Bell-Telephone Laboratories.
The CO maps are sampled at intervals of $\Delta b = 2'$ with a
half-power beamwidth of $100''$, and the latitude coverage
is $\simeq\pm 2^\circ$. The longitude coverage of the HI and CO data
are $l= 33^\circ\sim64^\circ$ and $l=10^\circ\sim77^\circ$, respectively.
Both maps are spaced at equal intervals in longitude:
$\Delta\sin l = 0.025$. We used the Arecibo HI data,
which have the highest resolution for HI observations, in order to
compare with the CO data of a resolution $100''$ and to examine thevertical
(latitude) variation of gas density. Table 1 shows the observational
parameters of the data, and Figure 1 shows examples of the HI and CO $b-v$ 
diagrams superposed on each other.

\begin{center} -- Table 1, Figure 1 -- \end{center}

\subsection{Molecular fraction}
We adopt the following relations between the column density $N$ of the
atomic and molecular hydrogen and the intensity $I$ of the HI and
CO line emissions:
	\begin{equation}
	N_{\rm H  }\; [\:{\rm cm^{-2}}] = \, C_1 I_{\rm HI} \:
	\end{equation}
and
	\begin{equation}
	N_{\rm H_2}\, [\:{\rm cm^{-2}}] = \, C_2 I_{\rm CO} \:,
	\end{equation}
where $C_1$ and $C_2$ are the conversion factors.
In this paper we regard $C$ as a constant and adopt
	\begin{equation}
	C_1 =1.8 \times 10^{18} \enskip
	\rm{cm^{-2}/(K\thinspace km\thinspace s^{-1}})
	\end{equation} 
and
	\begin{equation}
	C_2 =2.8 \times 10^{20} \enskip 
	\rm{cm^{-2}/(K\thinspace km\thinspace s^{-1}})
	\end{equation} 
(Burton 1988; Bloemen {\it et al.} 1985). However, Arimoto {\it et al.}
(1996) point out that the CO-to-H$_2$ conversion factor in spiral
galaxies is dependent on the radial distance from the Galactic Centeras
	\begin{equation}
        C_2=C_2^0 \;10^{0.39\:R/R_e},
	\end{equation} 
where the coefficients have been given as
$C_2^0 = 0.9 \times 10^{20} \enskip \rm{cm^{-2}/
(K\thinspace km\thinspace s^{-1}})$
and
$R_e= 6.6$ kpc.
Although our study will be based on an assumption of a constant
$C_2$ factor as equation (4), we examine the effect of the
radial variation of $C_2$ on our result, and discuss it in the
last section.

The number density $n$ is given by
	\begin{equation}
	n = N / \Delta r ,
	\end{equation}
where $\Delta r$ is the depth along the line of sight.

We define the molecular fraction $f_{\rm{mol}}$ as the ratio of the
molecular hydrogen gas density to the total hydrogen gas density,
which is expressed by mass density $\rho$ or number density $n$ by
	\begin{equation}
	f_{\rm{mol}}
               =\frac{\rho_{\rm{H_2}}}{\rho_{\rm{H}}+\rho_{\rm{H_2}}}
               =\frac{2\,n_{\rm{H_2}}}{n_{\rm{H}}+2\,n_{\rm{H_2}}}. 
	\end{equation}

\subsection{Tangent points and Terminal-velocity method}

Assuming a circularly symmetric model of the Galaxy, we can obtain
distances to the emission at terminal velocities in the first and
fourth quadrants ($0^\circ<l<90^\circ$ and $270^\circ<l<360^\circ$).
The radial velocity at each galactic longitude $l$ in the first quadrant
is given by
	\begin{equation}
	v_r = \left[  \frac{R_0}{R} V(R)-V_0 \right] \sin l ,
	\end{equation}
where $V(R)$ is the rotation velocity at galactocentric distance
$R$, $R_0(=8.5\;\rm{kpc})$ is the distance of the sun from the galactic
center, and $V_0(=220\;\rm{km\, s^{-1}})$ is the rotation velocity
at $R=R_0$.

Terminal-velocity emission comes from the tangent point at a
galactocentric radius
	\begin{equation} 
	R=R_0\,\vert \sin l \,\vert ,
	\end{equation}
and the galactic latitude $b$ corresponds to the height from the galactic
plane as
	\begin{equation} 
	z=R_0\,\cos l \,\tan b .
	\end{equation}

The actual velocity profiles do not have a sharp cutoff due to the velocity
dispersion of the gas. So, we regard the integrated emission within a
certain velocity range of width $\Delta v$ around the terminal velocity
as a contribution of the gas at the tangent point. We adopt a velocity
width $\Delta v = 6.5\thinspace {\rm km \thinspace s^{-1}}$ at any
longitude, which we found to be a typical width of the CO velocity
profile corresponding to a terminal-velocity emission component
at $b\simeq0$. In fact, this value is consistent with the velocity
dispersion obtained by Malhotra (1994), who finds that the value of
the velocity dispersion varies
3.8$\thinspace {\rm km \thinspace s^{-1}}$ at 2.5 kpc to 7.1
$\thinspace {\rm km \thinspace s^{-1}}$ at 7.5 kpc.
The depth along the line of sight corresponding to this velocity width is
given by
\begin{eqnarray}
\Delta r &=& 2R_0\,\sin l \;
             \left[\left(\frac{V_0}{V_0-\Delta v}\right)^2-1\right]^{1/2} \\
                      &\approx& 0.50 \,R_0 \,\sin l.\nonumber
\end{eqnarray}

Based on these relations we calculated number densities of
H and H$_2$ at each longitude and latitude using
latitude-velocity ($b-v$) diagrams, and obtained vertical
density distributions as a function of height $z$ from the
galactic plane. Then, we arranged the thus obtained $z$ distributions
at various longitudes in the order of the distance from the galactic
center, and obtained two dimensional maps of the distributions of
H and H$_2$ gases in the $(R,z)$ plane.

\section{Result}
\subsection{HI and H$_2$ Distributions}
As seen from the $b-v$ diagram (Figure 1), the HI gas is distributed
smoothly in the velocity direction at any longitude, and therefore,
at any point along the line of sight.
On the other hand, the H$_2$ gas distribution is clumpy.
The HI and H$_2$ distributions in the $z$ direction are also clearly
different (Figure 2): the HI distribution has a flat plateau near the
galactic plane and gentle skirts, while most of H$_2$ profiles have
multiple peaks, and the maximum density is much higher than that of HI.
The CO distribution is not centered on the Galactic plane, which is
consistent with previous studies (Malhotra 1994; Dame {\it et al.} 1987).

\begin{center} -- Figure 2 -- \end{center}

Figures 3{\em a} and {\em b} show the  $(R,z)$ distributions of HI
and H$_2$, respectively. These figures, therefore, represent cross
sections of our Galaxy along the tangent velocity circle.
The density distribution of HI gas in the $z$ direction extends for a
few hundred parsecs, while the extent of H$_2$ is not more than 100 pc:
there is little CO emission at high latitudes. Note, however, that these
two cross sections are not fully sampled in the longitude direction,
because the observations have been obtained every
$\Delta \sin l = 1/40\; (\Delta l = 1.^\circ 4\:/\cos l :\Delta R = 212.5\;
{\rm pc} )$ with a beam width $4'$ for HI and $100''$ for CO.

\begin{center} -- Figure 3 -- \end{center}

In Figure 4, we show an $(R,z)$ distribution of H$_2$ gas for the whole
range covered by the present CO data from the galactic center region
($l=10^\circ, R=1.49 \:$kpc) to the solar neighborhood
($l=77^\circ, R=8.29 \:$kpc). This figure includes the region shown in
Figure 3{\em b}. Figure 5($a$) shows radial distributions of HI and
H$_2$ gases averaged at  $\vert z \vert < 100 {\rm pc}$.
A radial distribution of CO emissivity in our Galaxy has been obtained by
Robinson {\it et al.} (1988) and Sanders {\it et al.} (1984).
Our result is approximately agree with theirs.
The variation of the H$_2$ number density is much larger than that of HI.
Note that the highest density region for H$_2$ in Figure 4 and 5
corresponds to the `4-kpc molecular ring', and that the H$_2$ gas density
in the inner region within 3$\:$kpc and outside of 7$\:$kpc is an order of
magnitude smaller than that in the 4-kpc ring.
Figure 5($b$) shows the same as Figure 5($a$) but using the $C_2$ factor
as given by equation (5), which takesinto account the radial variation of the CO-to-H$_2$ conversion factor.
The H$_2$ density distribution changes drastically in the inner region of
the Galaxy, indicating much less molecular gas within $R\sim 4$ kpc.

\begin{center} -- Figure 4, 5 -- \end{center}

\subsection{Variation of \bf $f_{\rm mol}$}
We calculated $f_{\rm mol}$ for a region shown in Figure 3,
where both the HI and H$_2$ densities are obtained.
Negative values in the observed CO intensity (H$_2$ density)
have been replaced with zero.
Figure 6 shows an example of the $z$-variation of $f_{\rm mol}$
at $l=53^\circ$. It is evident that the value of $f_{\rm mol}$ is
almost unity near the galactic plane, and decreases rapidly to zero in a
narrow range of $z$ as $z$ increases.

This phenomenon is commonly seen at all longitudes except at
$l=61^\circ$ and $64^\circ$ ($R\leq7.5$ kpc) where the amount of H$_2$ is
small. This fact indicates the existence of a $z$-directional boundary of
a region with a high H$_2$ fraction, which we call the molecular front in
the $z$-direction. Most of hydrogen gas is in the molecular phase in the
region near the galactic plane, sandwiched between the two molecular fronts,
whereas in the outer (higher) region it is in the atomic phase.

\begin{center} -- Figure 6 -- \end{center}

Figure 7 shows the two dimensional distribution of $f_{\rm mol}$ as a
contour map in the $(R,z)$ plane. The centroid of the molecular dominant
disk, where $f_{\rm mol}\geq0.8$, is waving, corresponding to the fact that
the molecular gas disk is corrugating (e.g. Malhotra 1994). It is remarkable
that the thickness of the molecular dominant disk is almost constant at
$109\pm12$ pc. On the other hand, it is known that the scale height of the
HI layer is larger than it. A thickness of HI layer between half-density
points has approximately constant value: 220 kpc over $4\leq R \leq 8$ kpc
 (Dickey and Lockman 1990). When we compare the thickness of molecular
 dominant disk with the scale height of HI gas, we can express that the
 atomic hydrogen layer wraps the molecular disk.

\begin{center} -- Figure 7 -- \end{center}

Figure 8($a$) shows radial variations of the molecular fraction calculated
for number densities of HI and H$_2$ averaged within $\vert z \vert <$ 50
and 200 pc of the galactic plane. In both cases the molecular fraction
decreases from 0.8 to 0.6 with increasing $R$. Though we can not clearly
see the phase transition in this region, curves are consistent with the
fact that the molecular fraction is almost unity in the the central few
kpc region (Sofue {\it et al.} 1994). Particularly,
the curve for $\vert z \vert <$ 200 pc indicates a decrease of
$f_{\rm mol}$ beyond $R\simeq$ 7.5 kpc.
This may be consistent with the existence of the radial molecular
front indicated from an analysis of longitude-velocity diagrams of the
Galaxy (Sofue {\it et al.} 1994). Since the H$_2$ gas density over
$R \geq 8$ kpc is much smaller than that of inner region
(Dame {\it et al.} 1987), it is expected that the molecular fraction
will decrease suddenly beyond this radius.
Figure 8($b$) shows the same variation but considering the radial
dependency of CO-to-H$_2$ conversion factor.
The radial molecular front slightly change.
However, no significant effect is found.
\begin{center} -- Figure 8 -- \end{center}
 
\subsection{Model Analysis}
The three crucial parameters which determine $f_{\rm mol}$ are the
interstellar pressure $(P)$, the intensity of dissociative UV radiation
field $(U)$, and the metal abundance $(Z)$ which is related to the
amount of interstellar dust and shields UV photon (Elmegreen 1993).
Based on the phase-transition theory (Elmegreen 1993),
Honma {\it et al.} (1995) has developed a computing code to model
the molecular front in galactic scale. By applying this code,
we calculate the variation of molecular fraction in the $z$-direction.

Since it is reasonable to assume that the pressure is proportional to the
gas density, we consider the following two cases: first, $P(z)$ is
traced by the observed total gas density distribution in the
$z$-direction ($n_{\rm tot}$), and second, $P(z)$ is expressed by a
gaussian function of $z$, considering the hydrostatic balance of the gas
pressure with the gravity in the $z$-direction.
According to van der Kruit and Searle (1982) we assume that the intensity
of the radiation field is expressed as
	\begin{equation}
	U(R,z) = U_0\; e^{(-R/R_{\rm U})} {\rm sech^2}(z/z_0) ,
	\end{equation}
where $R_{\rm U} = 5.0$ kpc is a scale distance from the galactic center,
and $z_0 = 0.7$ kpc is the scale height  in the $z$-direction.
For the metallicity, we do not have any information about the gradient
in the $z$-direction. So, we assume that the value of $Z$ is constant
against $z$, while it varies with $R$ as
$Z=Z_0\; e^{- \left(R-R_0 \right)/R_Z }$.

\begin{center} -- Figure 9 -- \end{center}

Figure 9 shows a thus calculated variation of the molecular fraction for
$l=53^\circ$.
The solid line corresponds to the observed $n_{\rm tot}(z)$,
and the dashed line to the gaussian distribution of $P(z)$ in which the
parameter of the gaussian function was taken so that it fits the
observed $n_{\rm tot}(z)$. Even for such a round pressure distribution
as a gaussian, $f_{\rm mol}$ shows a shoulder-like variation,
indicating the phase transition clearly. The curve of $f_{\rm mol}$
calculated for the observed $n_{\rm tot}(z)$ shows a sharper
variation at a critical height, and agrees well with the observed
$f_{\rm mol}$ variation (Figure 6) and reproduces the molecular front in
the $z$-direction. Though it has been considered that the most important
parameter for the molecular front is the metal gradient (Honma {\it et al.}
1995), our results shows that the molecular front is reproduced without
a metal gradient.

\section{Discussion}
The radial molecular front has been observed in our Galaxy as well as in 
many external galaxies (Sofue {\it et al.} 1994; Honma {\it et al.} 1995), 
in which the front is found at $R \sim 5-10\:{\rm kpc}$. In our result for 
our Galaxy, although $f_{\rm mol}$ is almost unity in $4.8\leq R \leq 7.2$ 
kpc, the radial decrease of $f_{\rm mol}$ may occur at $R\geq$ 7.5 kpc, 
where the value of $f_{\rm mol}$ drops from $\sim$ 0.8 to less than 0.6. 
Since the HI gas density varies more gradually than the H$_2$ gas 
at $R\geq7\;{\rm kpc}$, we may expect that the molecular gas is not any more 
dominant in the outer region of $R=7.5$ kpc. This indicates that the 
molecular front in our Galaxy is located at $R\simeq$ 7.5 kpc and the 
transition region from H$_2$ to HI and vice versa is as 
narrow as $\Delta R \sim 2\;{\rm kpc}$ in the radial direction.

From the $z$-directional variation of $f_{\rm mol}$, we found the molecular 
fronts in the direction perpendicular to the galactic plane. 
The molecular hydrogen disk is sandwiched by a thicker HI envelope. 
In the molecular region near the galactic plane, between the two fronts, the 
ISM is almost totally in the molecular phase with $f_{\rm mol} \geq 0.8$, 
while in the outer region, the gas is almost totally HI. We stress that the  
transition between HI and H$_2$ is taking place within the narrow range at a 
height of $z\simeq \pm 50\;$ pc at $R\leq7$ kpc. 

In this paper, we have not taken to account a metallicity gradient in 
the $z$-direction, since the existence of the $z$-gradient of the 
metallicity has not been
reported. Even if it exists, the scale height of the variation would be a few
hundred pc, because it will reflect the scale height of old population
stars. Therefore, it will be reasonable to assume 
that the metallicity gradient in
$z$-direction is small or not present in our region of $|z| < 100$ pc. 

It has been found that the CO-to-H$_2$ conversion factor
($C_2$ factor) increases exponentially with the radius due to the
metallicity gradient (Arimoto {\it et al}. 1996). This implies that
the $C_2$ value in the region considered here (from R=4 to 7 kpc)
varies by a factor of 1.5. We have examined the radial variations
of the H$_2$ density and $f_{\rm mol}$ by taking into account this effect.
We found that the radial molecular front slightly 
moves outwards, but not significant change of the property
of the $f_{\rm mol}$ distribution in the $z$-direction was found.

\bigskip
We thank M. Honma for his valuable comments and for providing us with the 
code to model the molecular front. We are also indebted to Drs. T. M. 
Bania, G. R. Knapp, and S. Malhotra for providing us with the HI and CO 
data in a machine readable formats.

\bigskip\bigskip
\newpage\normalsize

\begin{description}
\item[]{\normalsize\bf References}
\item[]Arimoto N., Sofue Y., and Tsujimoto T., 1996, PASJ Vol 48 in press.
\item[]Bania, T. M., and Lockman, F. J., 1984, ApJS 54, 513.
	% HI Survey of Our Galaxy
\item[]Bloemen, J.B.G.M., Strong, A. W., Cohen, R. S., Dame, T. M., 
Grabelsky, D. A., Hermsen, W., Lebrun, F., Mayer-Hasselwander, H. A., 
and Thaddeus, P., 1985, A\&A 154, 25. 
	%conversion factor 
\item[]Burton, W. B., and  Gordon, M. A., 1978, A\&A 63, 7.
\item[]Burton, W.B. 1988 in  Galactic and Extragalactic Radio Astronomy, 
edited by G. L. Verschuur and K. I. Kellermann (Springer-Verlag, Berlin), 
Chap. 7. %HI in MW, review
\item[]Combes F., 1992, ARA\&A 29, 195.%CO in MW, review 
	% Distribution of CO in The Milkey Way
\item[]Dame, T. M., Ungerechts, H., Cohen, R. S., de Geus, E. J., 
Grenier, I. A., May, J., Murphy, D. C., Nyman, L. -A., and Thaddeus, P., 
1987, ApJ 322, 706.
	% CO Survey of Milkey Way
\item[]Elmegreen, B. G., 1993, ApJ 411, 170.
	% The HI and H2 Transition in Galaxy
\item[]Honma, M., Sofue, Y., and Arimoto, N., 1995, A\&A 304, 1.
	% Radial Variation of the Molecular Fraction in Galaxies
\item[]Knapp, G. R., Stark, A. A., and Wilson, R. W., 1985, AJ 90, 254.
	% CO Survey of Our Galaxy
\item[]Lockman, F. J., 1984, ApJ 283, 90.
	% The HI Halo in the Inner Galaxy
\item[]Malhotra, S., 1994, ApJ 433, 687.
	% The Vertival Equiliblium of Molecular Gas in the Galactic Disk
\item[]Sanders, D. B., Solomon, P. M., and Scouville, N. Z., 1984, 
ApJ 276, 182.
	% Giant Molecular Clouds in the Galaxy
\item[]Sofue, Y., Honma, M., and Arimoto, N., 1995, A\&A 296,33.
%\item[]Sofue, Y., and Tosa,M., 1974, A\&A 36, 237.
%\item[]Solomon, P. M., Rivolo, A. R., Barrett, J., and Yahill, A., 1987, 
ApJ 319, 730 %conversion factor in the Galaxy
\item[]van der Kruit, P.C., and Searle, L., 1982, A\&A 110, 61.
\end{description}
\newpage

\begin{description}
\item[{\normalsize\bf Figure captions}]
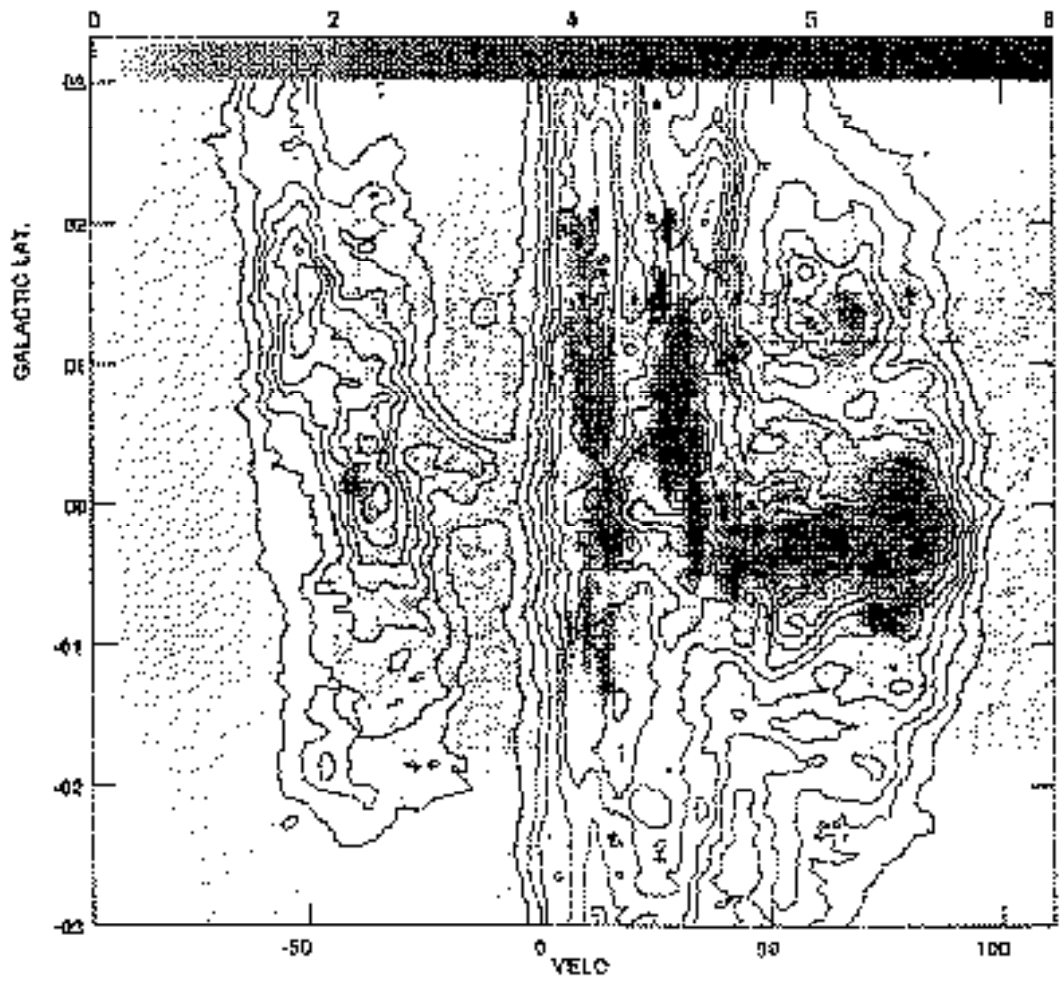
\item[Figure 1.]
A latitude-velocity ($b-v$) contour map of the HI brightness temperature 
at $l=39^\circ$. The ordinate is galactic latitude in degrees, 
and the abscissa is velocity relative to the LSR in ${\rm km \: s^{-1}}$. 
Contours are drawn at 10\% interval of the peak brightness (114 K). 
A $b-v$ map of the CO emission at $l=39^\circ$ is superposed in gray scale. 
 
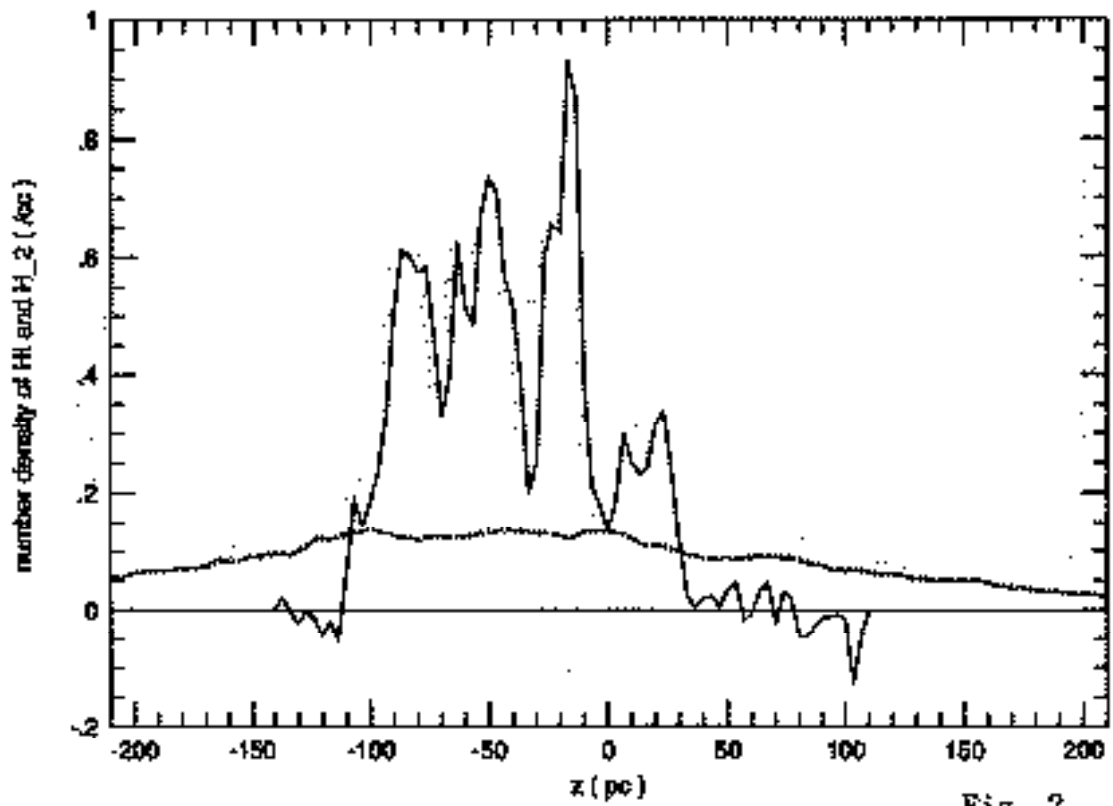
\item[Figure 2.]
Vertical distributions of the atomic hydrogen gas (doted line) and the 
molecular hydrogen gas (solid line) at $l=53^\circ$. The ordinate is the 
number density in HI ${\rm cm^{-3}}$ and H$_2$ ${\rm cm^{-3}}$. 
(Note that the ordinate scaling for H$_2$ should be doubled when 
comparing the two curves in mass.) The abscissa is the height from the 
galactic plane. 

\item[Figure 3.]
({\em a}) The $(R,z)$ distribution of the HI gas density. Contours are 
drawn at intervals of 0.01 H cm$^{-3}$. ({\em b}) The $(R,z)$ distribution 
of the H$_2$ gas density. Contours are drawn at intervals 
of 0.1 H$_2$ cm$^{-3}$.

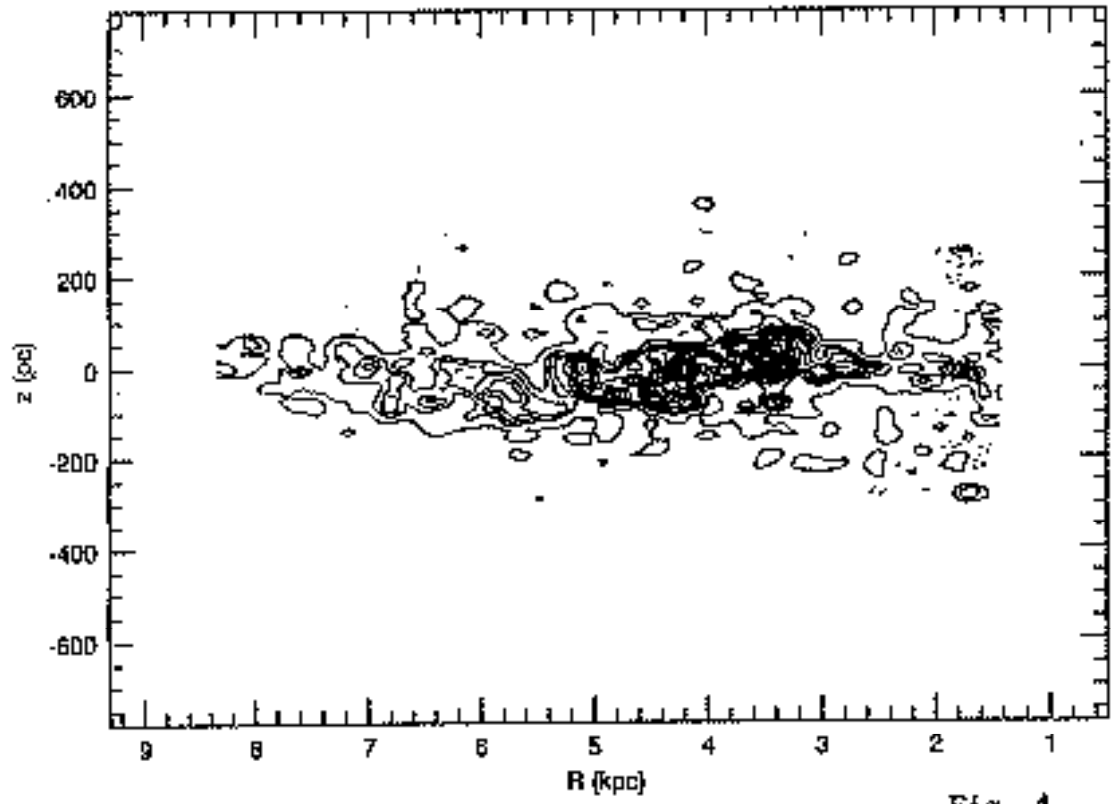
\item[Figure 4.]
The $(R,z)$ distribution of the H$_2$ gas density for the whole range 
of the CO data. Contours are drawn at intervals of 0.2 H$_2$ cm$^{-3}$.

\item[Figure 5.]
({\em a})Radial distributions of HI gas density (doted line) and H$_2$ gas 
density (solid line) averaged in $\vert z \vert < 100\; {\rm pc}$. ({\em b})
The same, but including the radial variation of the CO-to-H$_2$ conversion 
factor $C_2$ as given by equation (5).

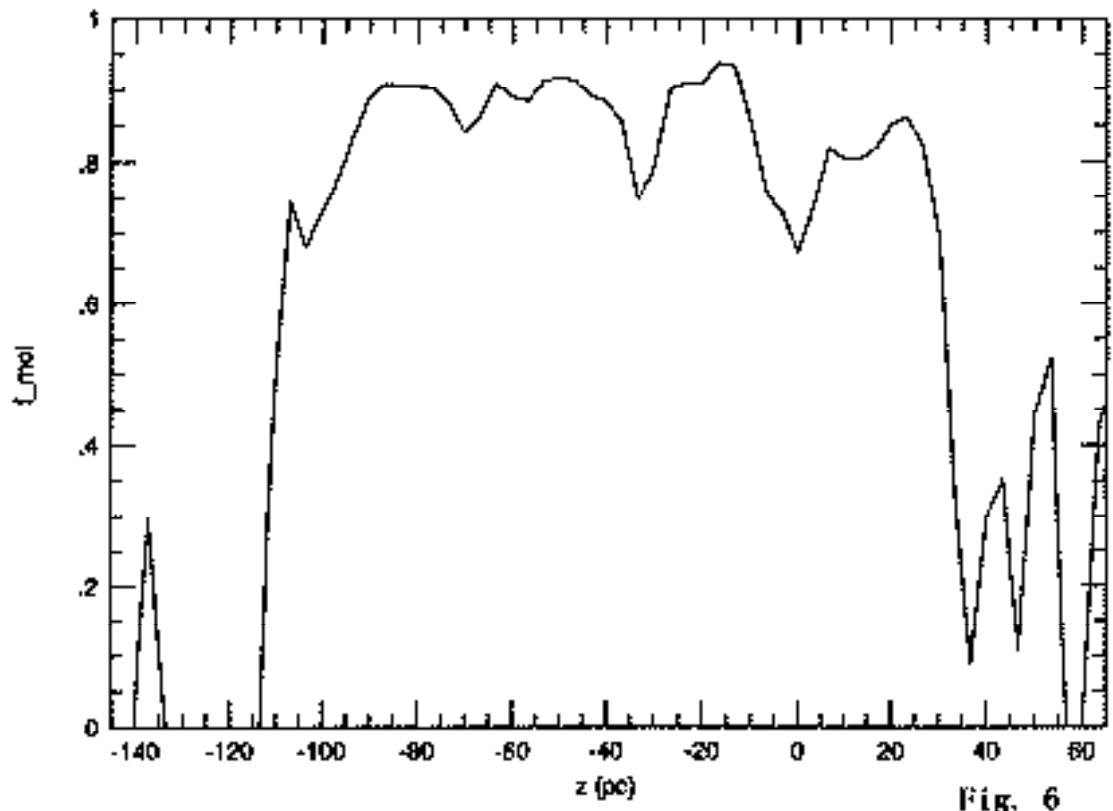
\item[Figure 6.]
The molecular fraction at $l=53^\circ$ as a function of $z$. 
The ordinate is $f_{\rm{mol}}$.

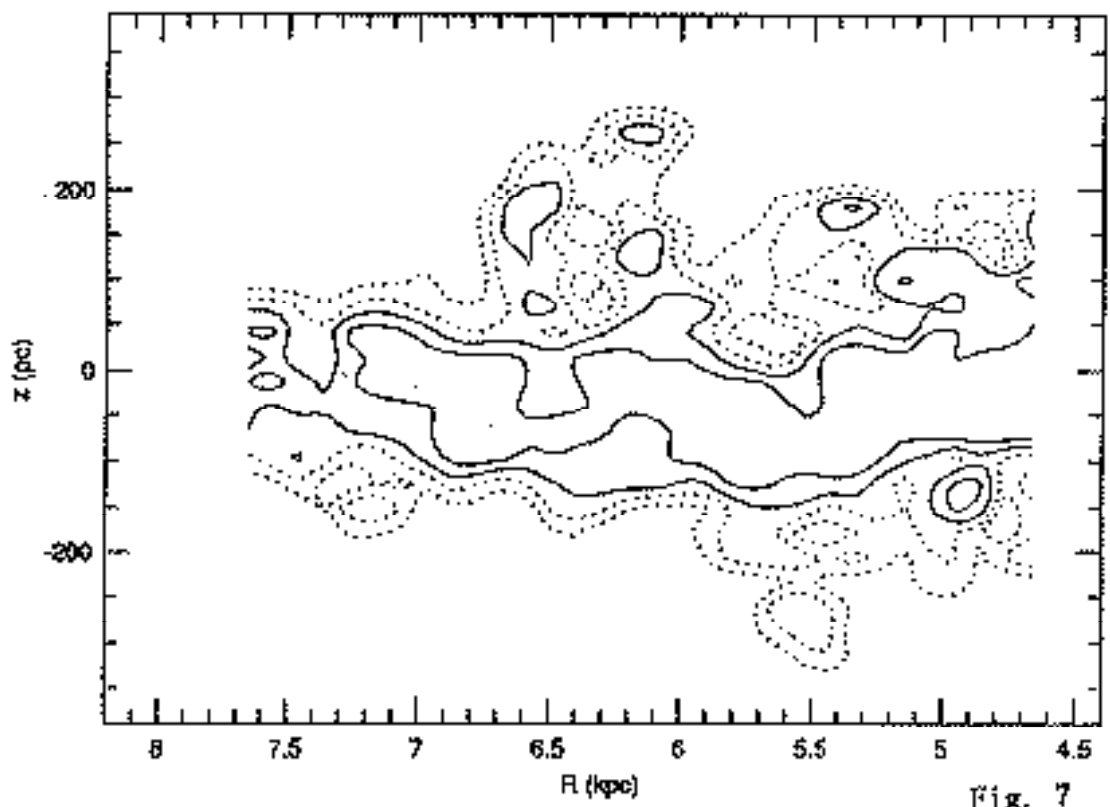
\item[Figure 7.]
The $(R,z)$ distribution of the molecular fraction. Contours are drawn 
at an interval of 0.2. Dashed lines show $f_{\rm mol}$= 0.2, 0.4, 
and solid lines $f_{\rm mol}$= 0.6, 0.8.

\item[Figure 8.]
({\em a}) Radial variations of the molecular fraction averaged 
within $\vert z \vert <$ 50 pc (solid line) and 200 pc (dashed line). 
({\em b}) The same as ({\em a}) but including the effect of radial 
dependency of CO-to-H$_2$ conversion factor.
\item[Figure 9.]
The molecular fraction calculated theoretically for $l=53^\circ$. 
The solid line is for the case where the observed $n_{\rm tot}(z)$ was 
used to calculate $P(z)$. The dashed line corresponds to a gaussian 
distribution of $P(z)$. The doted line shows the observed $f_{\rm mol}$ 
(Fig. 6).

\end{description}
\newpage

\smallskip
\begin{center}
   \begin{tabular}{l|c c}
   \multicolumn{3}{c}{Table 1. Observational parameters}\\
   \hline
   \hline
    & HI & CO ($J=1\rightarrow 0$) \\
    & Bania and Lockman (1984) & Knapp {\it et al.} (1985) \\
   \hline
    Telescope &  Arecibo &  Bell Laboratories \\
    Aperture & 304\,m & 7\,m \\
    Wavelength & 21\,cm & 2.6\,mm \\
    HPBW &  $4'$ & $1'40''$ \\
    Longitude extent & $33^\circ\sim64^\circ$ & $10^\circ\sim77^\circ$\\
    Latitude extent & $\pm 3^\circ$ & $\simeq\pm 2^\circ$ \\
    $\Delta v$ & $1.03{\rm \;km \: s^{-1}}$ & $0.65{\rm \;km \: s^{-1}}$ \\
    $\Delta b$ & $2'$ & $2'$ \\
    $R$ (kpc)* & 4.68 $\sim$ 7.65 & 1.49 $\sim$ 8.29 \\ 
   \hline
   
    Longitude interval & 
    \multicolumn{2}{c}{$\Delta\sin l = 0.025,\; 
                        {\rm or}\;\Delta l = 1^\circ\!.43/\cos l$}\\
    Radius interval    & 
    \multicolumn{2}{c}{$\Delta R = R_0\Delta\sin l = 212.5\; {\rm pc}$}\\
   \hline
   \multicolumn{3}{l}{* {\small The distance to the Galactic Center is taken to be $R_0=8.5\:{\rm kpc}$.}}\\
   \end{tabular}
\end{center}
\smallskip
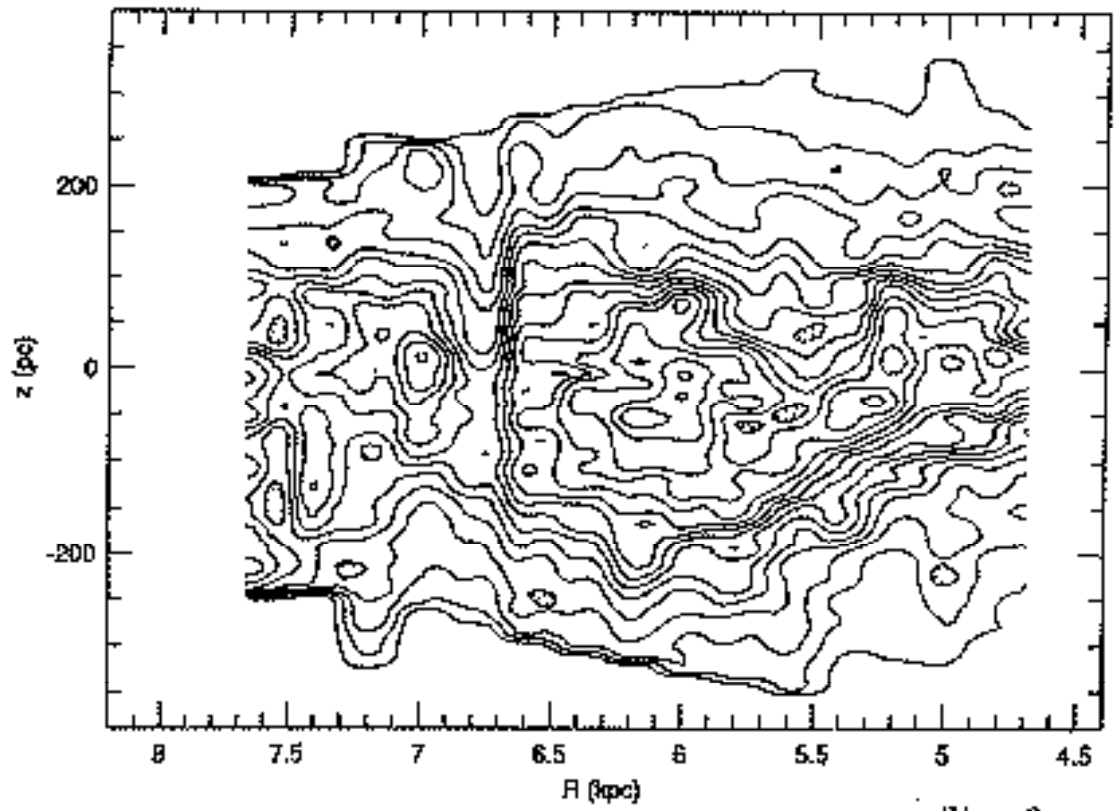
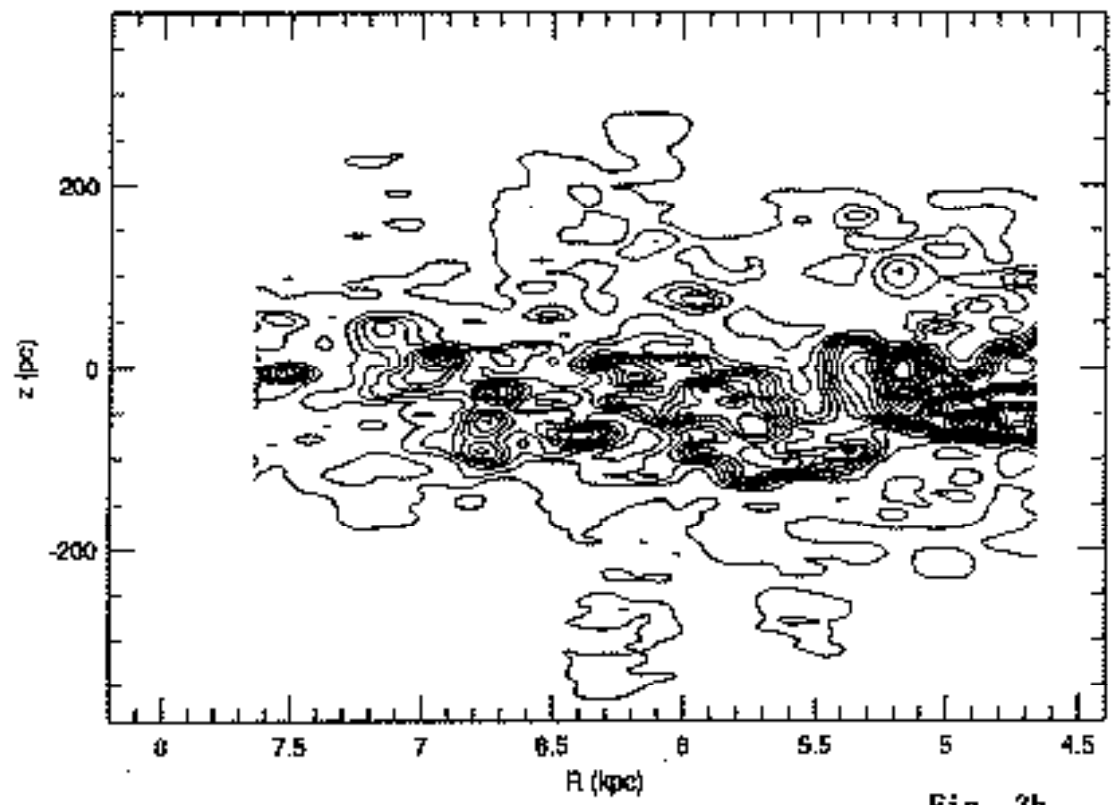
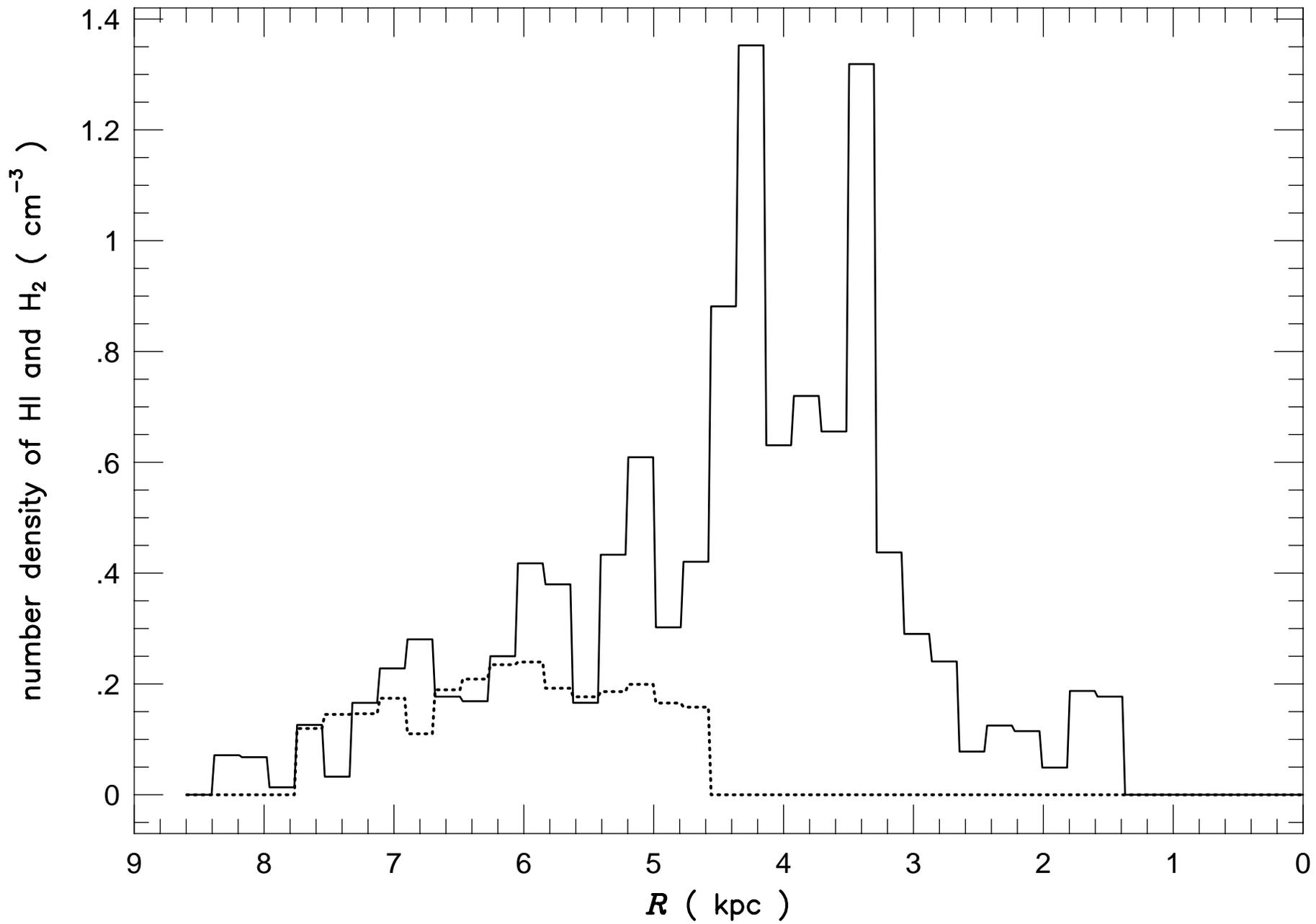
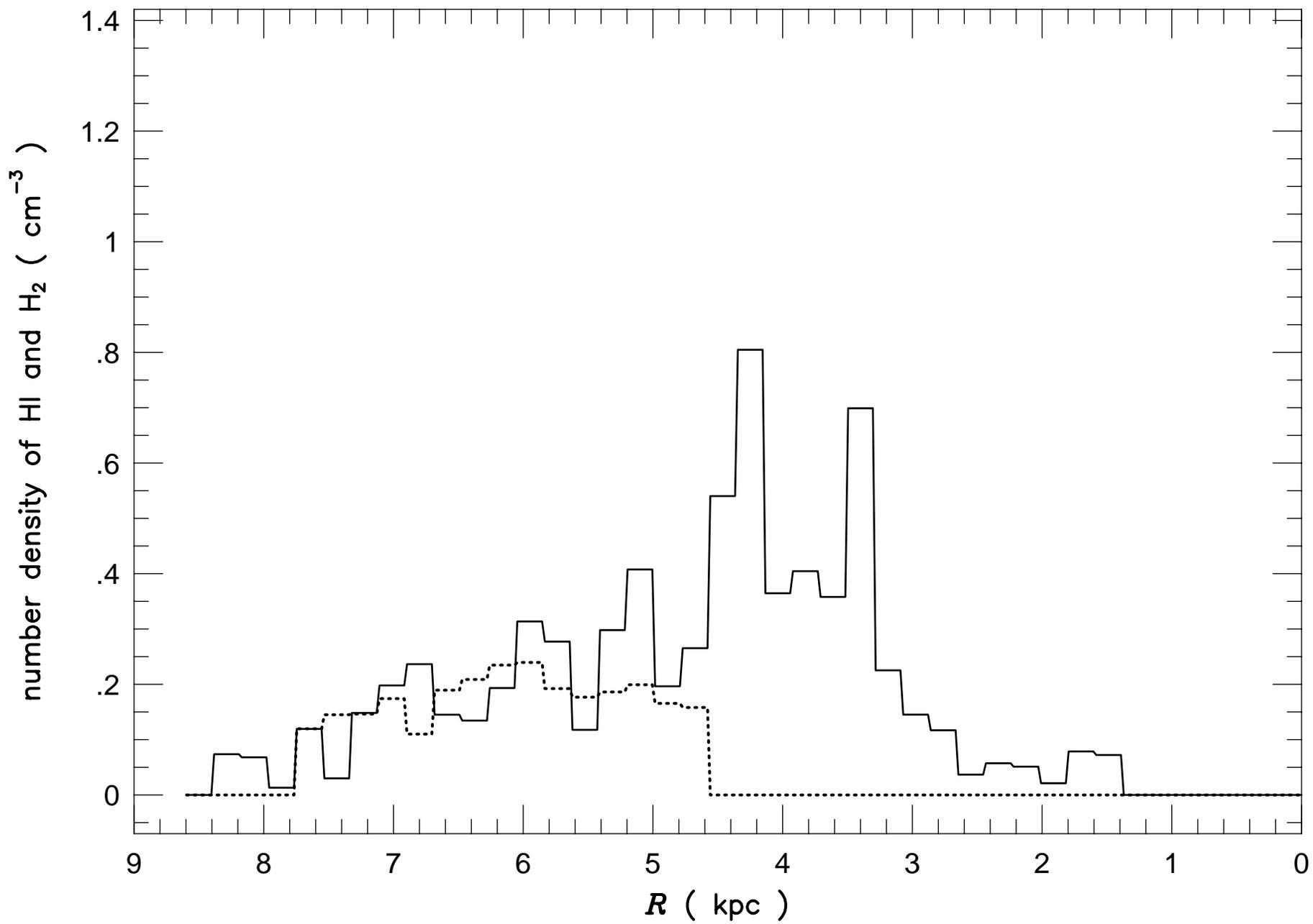
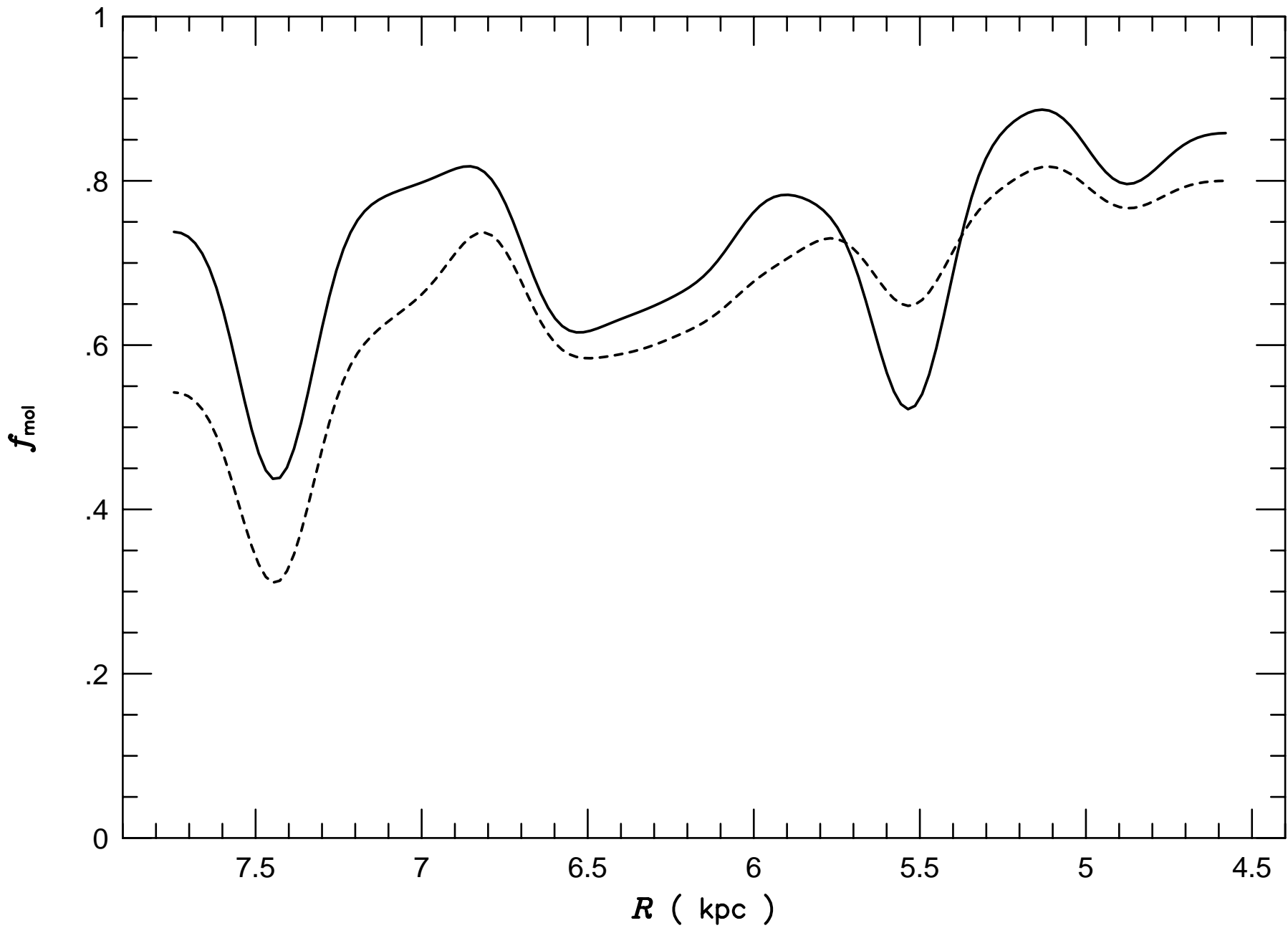
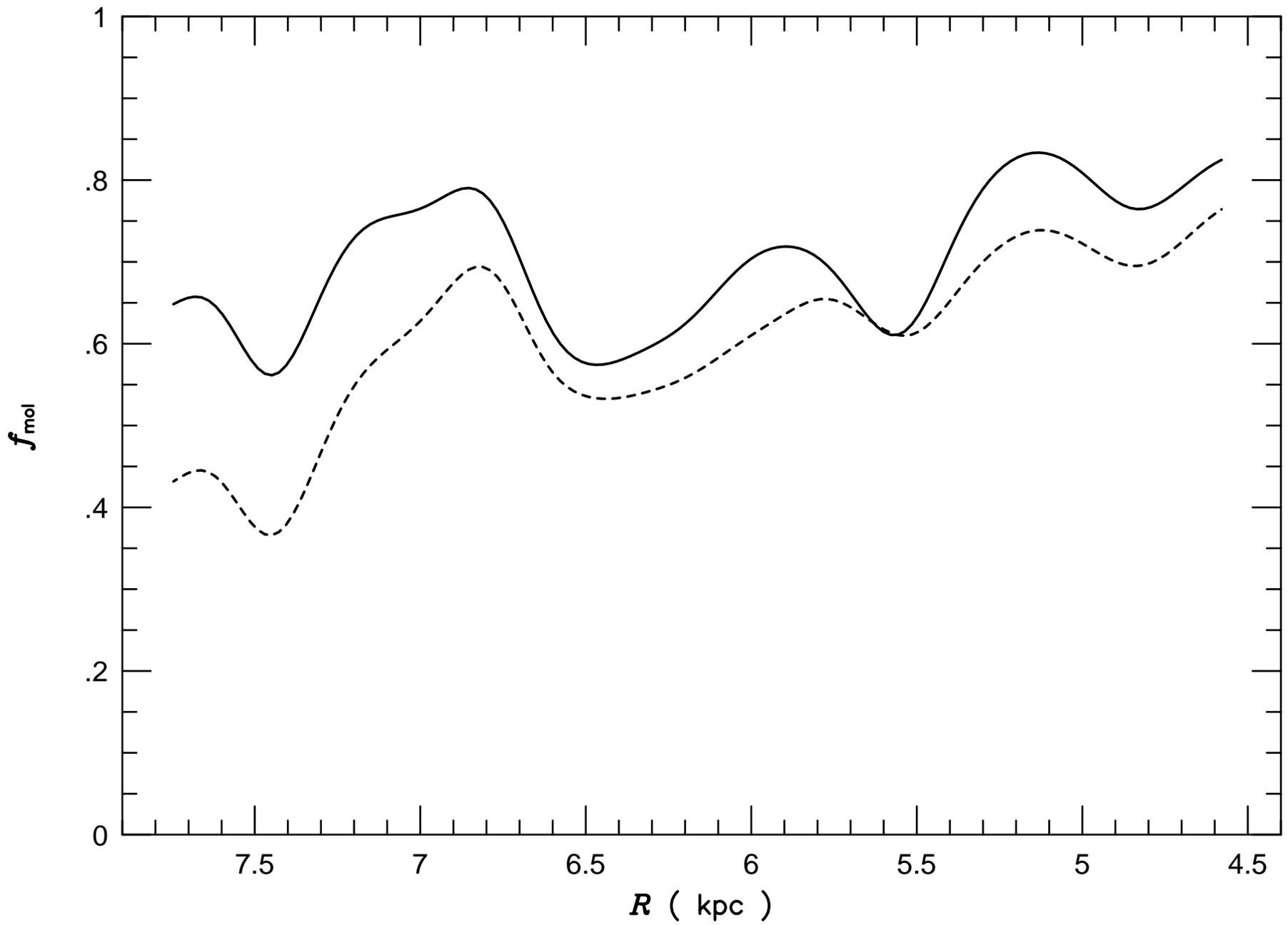
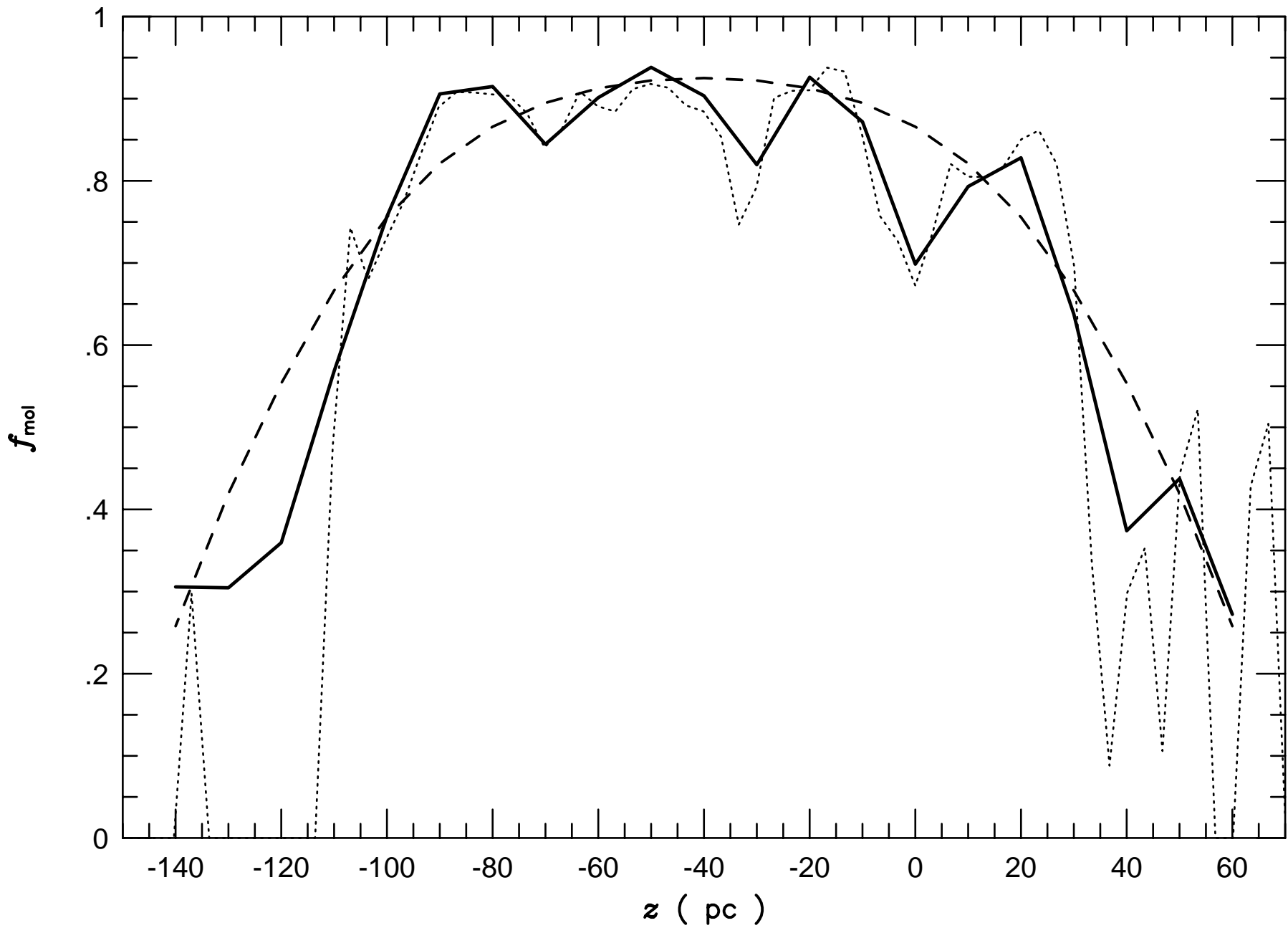
\end{document}